\documentclass[superscriptaddress,twocolumn,showpacs]{revtex4}
\usepackage[ansinew]{inputenc}
\usepackage [dvips]{graphicx}
\begin{document}
\title{Dependence of Curie Temperature on the Thickness of Epitaxial (Ga,Mn)As Film}

\author{B. S. S\o rensen}
\affiliation{Niels Bohr Institute fAFG, \O rsted Laboratory,
University of Copenhagen, Universitetsparken 5. DK-2100
Copenhagen, Denmark.}

\author{J. Sadowski}
\affiliation{Niels Bohr Institute fAFG, \O rsted Laboratory,
University of Copenhagen, Universitetsparken 5. DK-2100
Copenhagen, Denmark.} \affiliation{MAX-Lab, Lund University,
SE-221 00 Lund, Sweden.} \affiliation{Institute of Physics, Polish
Academy of Sciences, al. Lotnik\`{o}w 32/46, PL-02-668 Warszawa,
Poland.}

\author{S. E. Andresen}
\affiliation{Niels Bohr Institute fAFG, \O rsted Laboratory,
University of Copenhagen, Universitetsparken 5. DK-2100
Copenhagen, Denmark.}

\author{P. E. Lindelof}
\affiliation{Niels Bohr Institute fAFG, \O rsted Laboratory,
University of Copenhagen, Universitetsparken 5. DK-2100
Copenhagen, Denmark.}


\begin{abstract}
We present the magnetotransport properties of very thin (5 to 15 nm) single (Ga,Mn)As layers grown by low
temperature molecular beam epitaxy. A lower (Ga,Mn)As thickness limit of 5 nm for the ferromagnetic phase and the
dependence of the Curie temperature on (Ga,Mn)As thickness are determined from electrical transport measurements.
The Curie temperature is determined to be 97 K for the thinnest ferromagnetic sample and is found to decrease for
increasing layer thickness. A carrier density of ~7.1$\times$10$^{20}$ cm$^{-3}$ for the 5 nm thick (Ga,Mn)As layer
is determined from Hall measurements. Differences between magnetotransport properties of thick and thin (Ga,Mn)As
layers are observed and discussed.
\end{abstract}

\maketitle Since the discovery of ferromagnetism in III-V diluted magnetic semiconductors (DMS)~\cite{a,b}, they
have been proposed to be good candidates for model devices in the rapidly expanding field of Spintronics~\cite{c}.
(Ga,Mn)As is a ferromagnetic (FM) semiconductor and it is by now well established that ferromagnetism in this
compound is induced by exchange interactions between localised Mn magnetic moments and valence band
holes~\cite{d,e}. The highest Curie temperature of para- to ferromagnetic phase transition (T$_{c}$) obtained so
far for (Ga,Mn)As was known to be 110 K~\cite{f}, however slightly higher values, close to 130 K have been reported
recently by two different groups~\cite{g,h}.

As shown by many groups the electric and magnetic properties of (Ga,Mn)As are strongly dependent on the growth
conditions~\cite{i,j} and post-growth annealing procedures~\cite{k,l,m,n}, since in the case of (Ga,Mn)As with the
same Mn content, both factors are affecting concentration and/or distribution  of compensating donor-like defects
such as Mn interstitials~\cite{p,r} and As antisites~\cite{s,t}.

GaAs/(Ga,Mn)As Superlattices (SL) are promising candidates for enhancing T$_{c}$ due to the strong confinement of
holes in the magnetic layers in these structures~\cite{u}. In this context, the lower thickness limit of the
magnetic layer is interesting. In the studies of SL with magnetic (Ga,Mn)As layers separated by nonmagnetic AlGaAs
or InGaAs spacers, the ferromagnetic phase is seen to disappear when the thickness of the magnetic layers is below
5 nm~\cite{j,v,x}. Our recent investigations show that the paramagnetic- to ferromagnetic phase transition occur
for (Ga,Mn)As layers as thin as 4 - 8 ML (1.1 - 2.3 nm) in (Ga,Mn)As/GaAs SL~\cite{y,aa}.

The Curie temperature dependence on the thickness of single (Ga,Mn)As layers has previously been studied by F.
Matsukura \emph{et al.}~\cite{bb}. The studied samples had thicknesses in the range between 10-1000 nm, showing a
decreasing T$_{c}$ for increasing layer thickness, but no investigations on the lower limit for the ferromagnetic
phase to appear is reported.

Theoretical results concerning ferromagnetism  in (Ga,Mn)As quantum well structures~\cite{cc,dd} predict dependence
of Curie temperature on (Ga,Mn)As well width as well as the limit of thickness for the FM phase to occur to be
about 2.5 nm. According to the theoretical results of M. A. Boselli \emph{et al.}~\cite{cc} the FM phase transition
temperature in a (Ga,Mn)As quantum well is decreasing with increasing well width. It's therefore interesting to
investigate the low thickness limit for the FM phase to occur in (Ga,Mn)As in view of increasing T$_{c}$ beyond the
present limit of 110 K~\cite{f} and 130 K~\cite{g,h}.

Recently the FM phase transition has been detected by transmission magnetic circular dichroism measurements for
(Ga,Mn)As layers as thin as 2 nm~\cite{ee}, however to our knowledge no results of transport or magnetization
measurements have been published for single (Ga,Mn)As layers thinner than 10 nm.

In this letter we report that the experimental thickness for FM phase in single Ga$_{0.95}$Mn$_{0.05}$As layer
estimated from magnetotransport measurements is equal to 5 nm and the corresponding Curie temperature is T$_{c}$=97
K. T$_{c}$ is found to decrease with increasing film thickness. A carrier density of ~7.1$\times$10$^{20}$
cm$^{-3}$ is determined from Hall-measurements for the 5 nm thick layer at T=300 mK.
\newline

We have studied four samples grown by MBE under the same conditions and with the following growth sequence: 200 nm
standard high temperature GaAs buffer layer and 20 nm low temperature (LT) GaAs is grown on semi-insulating
epi-ready GaAs(100) wafers followed by a Ga$_{0.95}$Mn$_{0.05}$As layer with thicknesses of 3, 5, 10 and 15 nm
respectively. The (Ga,Mn)As layers are then capped by a 3 nm thick low temperature GaAs layer. The samples are made
in a 100 $\mu$m wide Hallbar geometry defined by standard UV-lithography and wet chemical etching. Au/Zn/Au
contacts are deposited, followed by annealing at ~240 ºC for one hour in a nitrogen atmosphere in order to create
ohmic contacts and to reach the optimum T$_{c}$~\cite{l,m,n}. All the data presented is obtained by standard
dc-measurements.

\begin{figure}[e]
\includegraphics[width=.45\textwidth]{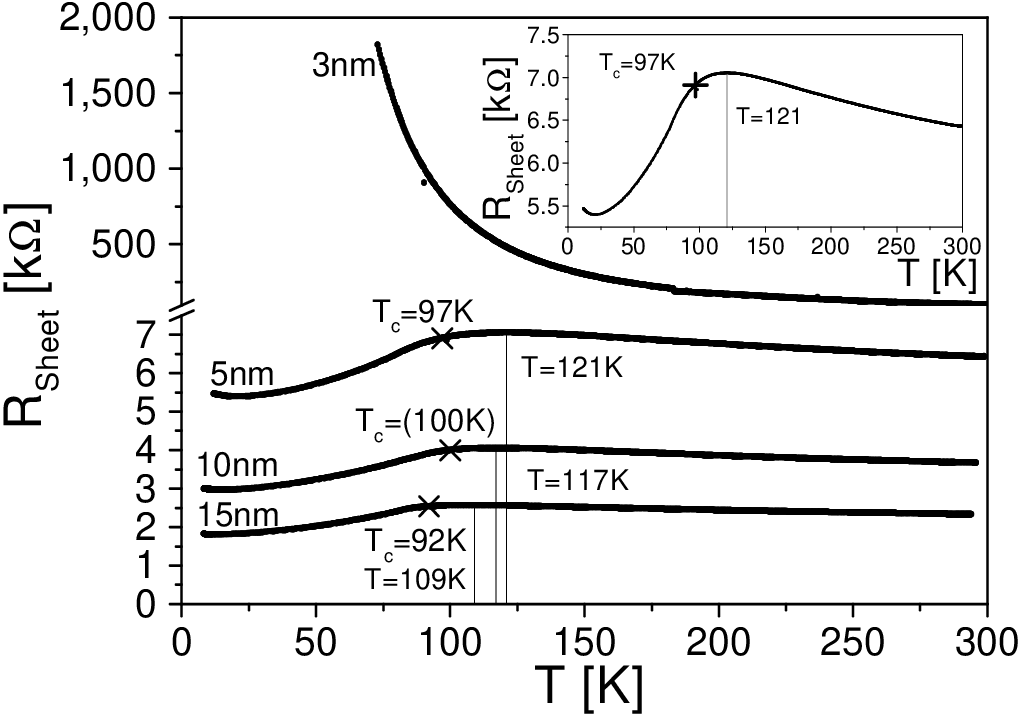}
\caption{\label{1}\ Sheet resistance $R_{sheet}$ as a function of temperature for four different thicknesses of
Ga$_{0.95}$Mn$_{0.05}$As thin film. The vertical lines for the different samples show the maximum in $R_{sheet}$.
The crosses shows the Curie temperature T$_{c}$, obtained from Hall measurements. An enlarged graph for the 5 nm
thin film is shown in the inset.}
\end{figure}

Figure 1 shows the sheet resistance of the four samples as a function of temperature, in zero magnetic field. The 3
nm sample shows an insulating behavior while the thicker samples have so-called metallic behavior, with a local
maximum in the resistance. This local maximum is common in magnetic metals and semiconductors and is in the region
of the transition temperature between the para- and ferromagnetic phase~\cite{d,l,m,y}. The maxima in the sheet
resistance for the samples with thickness of 5, 10 and 15 nm are indicated by the vertical lines and occur at
around T=121 K, 117 K and 109 K respectively, indicating that the Curie temperature decreases with increasing film
thickness. This is in agrement with the results by F. Matsukura \emph{et al.}~\cite{bb}. These data indicate a very
high Curie temperature compared with other thin films~\cite{bb,cc} and SL structures~\cite{j,m,y,aa}. The sheet
resistance of the 5, 10 and 15 nm thick samples decreases with increasing layer thickness as expected, giving a
resistivity of approximately 3.5 m$\Omega$cm at T=300 K.

\begin{figure}
\includegraphics[width=.45\textwidth]{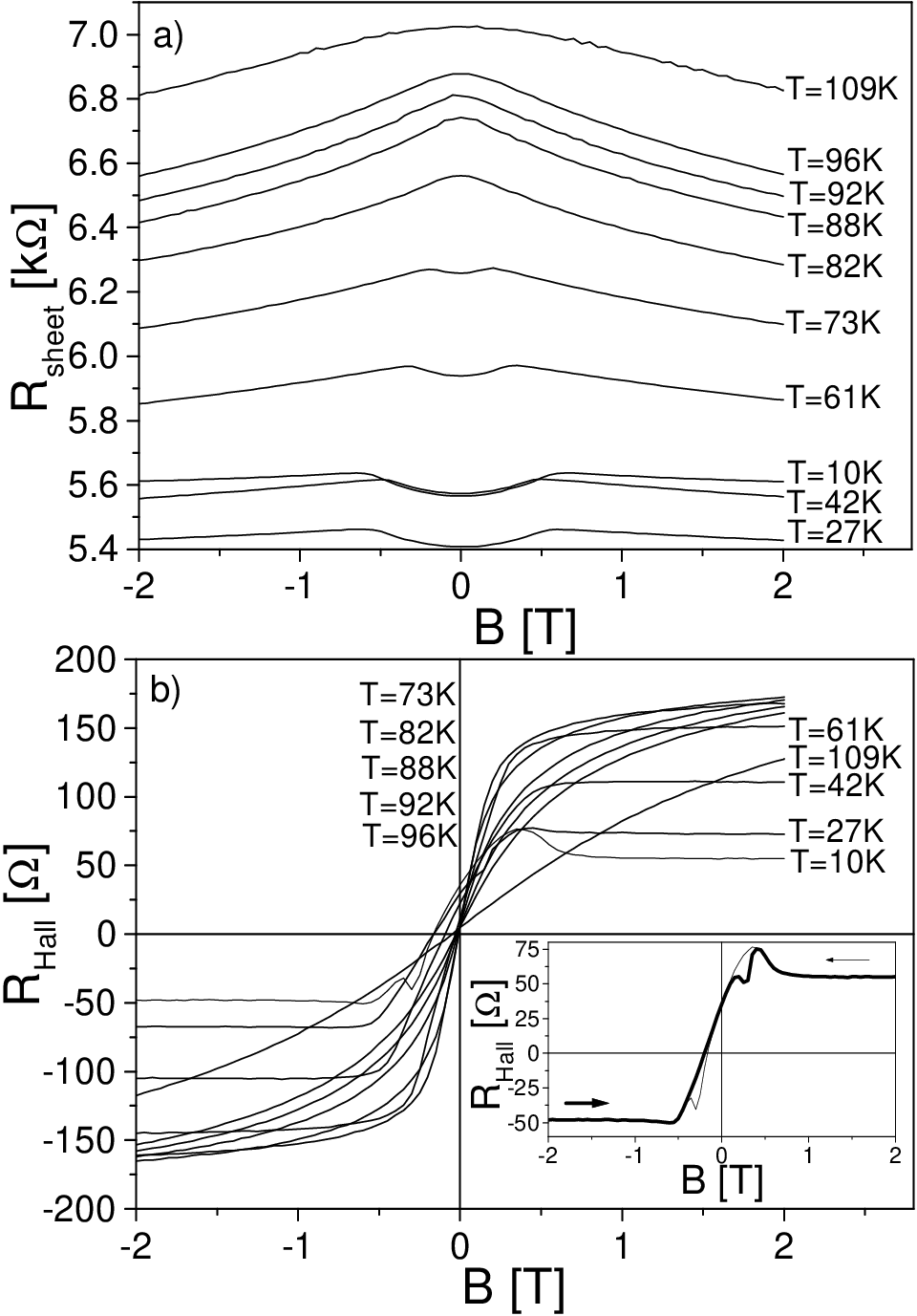}
\caption{\label{2}\ a) Sheet resistance $R_{sheet}$ as a function of perpendicular applied magnetic field for the 5
nm thick Ga$_{0.95}$Mn$_{0.05}$As thin film at different temperatures. b) The corresponding Hall resistance for the
same sample. The Hall resistance for different sweep directions are shown as inset.}
\end{figure}

An estimate of the Curie temperature has been obtained by Hall measurements in magnetic fields up to 2 T. The
sheet- and Hall resistance as a function of applied perpendicular magnetic field at different temperatures, is
presented in figure 2 for the 5 nm thick Ga$_{0.95}$Mn$_{0.05}$As thin film. The sheet resistance exhibit the same
behavior as seen in thick (Ga,Mn)As films~\cite{f}. The overall negative magnetoresistance can be understood as a
reduction of random spin-flip scattering due to the alignment of spins by the applied magnetic field. This has been
described for thick (Ga,Mn)As samples, at temperatures above T$_{c}$, by the spin disorder formula, as a decrease
in scattering of charge carriers by spin-fluctuations via the exchange coupling with increasing magnetic
field~\cite{f}. At temperatures below the Curie temperature a positive magnetoresistance appears at low magnetic
fields. The easy axis of magnetization is in the plane of the thin film. The positive magnetoresistance is believed
to be caused by the turning of the magnetization direction from the spontaneous in-plane direction to the
perpendicular direction by the applied magnetic field~\cite{f}, causing an increased spin-disorder. This behavior
of the magnetoresistance is seen in all the FM thin films.

\begin{figure}
\includegraphics[width=.45\textwidth]{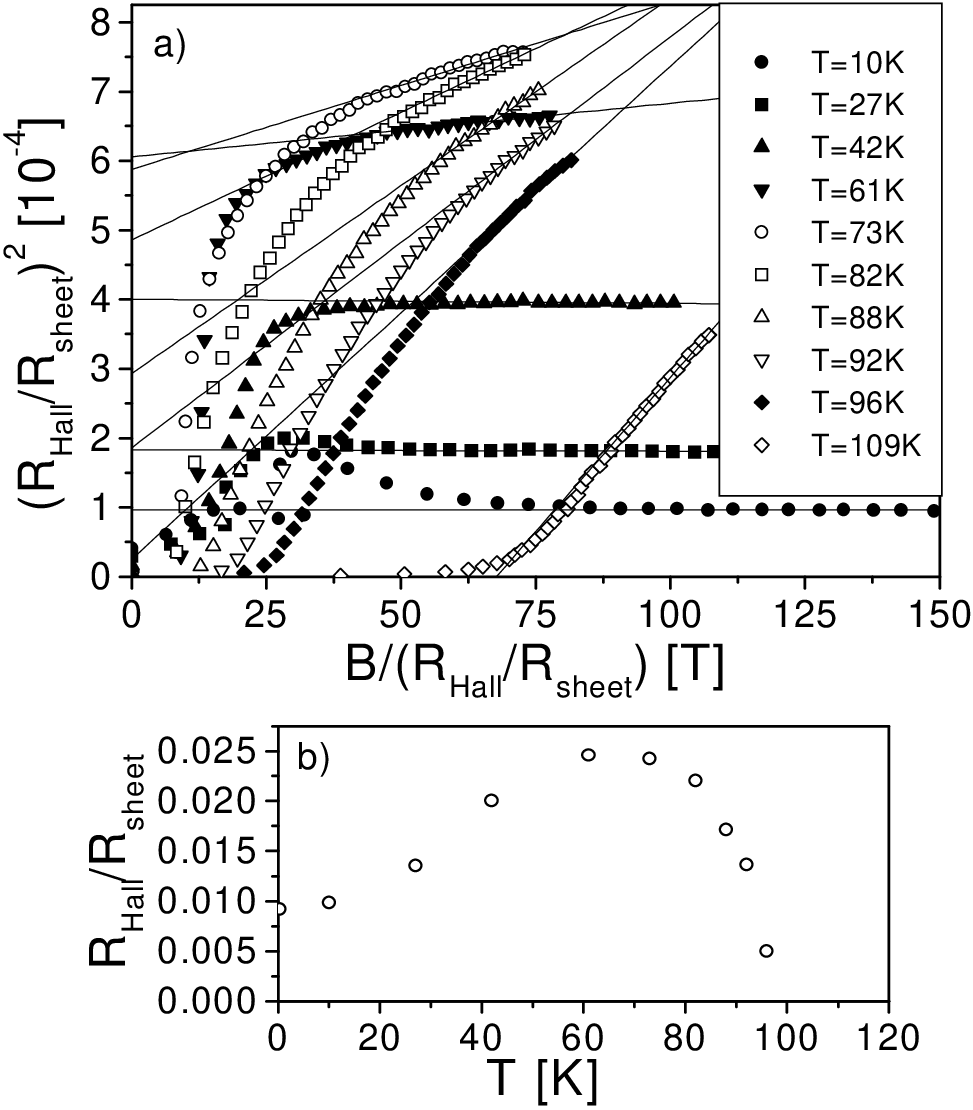}
\caption{\label{3}\ a) Arrott plot of the data obtained from the 5 nm thick Ga$_{0.95}$Mn$_{0.05}$As thin film at
different temperatures. The lines are linear extrapolations of the data at the highest magnetic fields. b) The
value of $R_{Hall}/R_{sheet}$ at the intersection as determined from the Arrott plot.}
\end{figure}

\indent The Hall resistance is dominated by the anomalous Hall effect, which may be written as:
\begin{equation}
R_{H}=\frac{R_{o}}{d}B+\frac{R_{S}}{d}M
\end{equation}
$R_{o}$ is the ordinary Hall coefficient and depends on the carrier density, $R_{S}$ is the anomalous Hall
coefficient and $M$ is the magnetization of the sample parallel to the applied magnetic field $B$. $d$ is the
thickness of the (Ga,Mn)As film. As can be seen in figure 2b, the Hall resistance at 2 T increases with decreasing
temperature. At 70 K it starts to decrease again. This behavior is seen in all the FM thin films. The Hall
resistance measured at 10 K is shown as an inset in figure 2b for both sweep directions of the magnetic field.
There is a broad bump at positive magnetic fields independent of sweep direction and a smaller anomaly that depends
on sweep direction. The broad bump does not appear in the thicker samples and the origin of this may be due to a
mixture of longitudinal and transverse resistances. The smaller anomaly is only present when the magnetic field is
swept from zero and out and appears before the magnetization is turned out of plane.

For thick (Ga,Mn)As layers, $R_{S}$ has been shown to be given by $R_{S}/d=cR_{sheet}$ (Skew
scattering)~\cite{b,f}. $c$ is a constant. Recent experiments shows that an additional term due to side-jump
scattering can be of importance~\cite{g}. Since the Anomalous Hall effect is dominant in this low magnetic field
range, it is not possible to determine the carrier density. The temperature at which the sample magnetization is
finite at zero applied magnetic field, can be found from extrapolation of a so-called Arrott plot assuming skew
scattering~\cite{f}. The Arrott plot for the 5 nm sample is shown in figure 3a, where $(R_{H}/R_{sheet})^{2}$ is
plotted as a function of $B/ (R_{H}/R_{sheet})$. The sample magnetization squared is proportional to the value of
intersection of the $(R_{H}/R_{sheet})^{2}$ axis by the back-extrapolated linear part from high magnetic fields.
From this, a Curie temperature of 97 K is determined. It should be noted that this is a low estimate of the Curie
temperature since a more accurate extrapolation can be obtained by applying a higher magnetic field. The Curie
temperature of the three samples determined from Hall measurements is shown in figure 1, together with the results
obtained from the maximum in the sheet resistance as a function of temperature. The discrepancy between the results
from the two methods is enhanced by the fact that T$_{c}$ determined by Hall measurement is a lower estimate, as
mentioned above, but it is also clear that the maximum in sheet resistance as a function of temperature is not a
good estimate of the Curie temperature in these thin films. There is some uncertainty as to T$_{c}$ for the 10 nm
thick sample, since the sample changed during Hall measurement.

The value of $R_{Hall}/R_{sheet}$ at the intersection from the Arrott plot is shown in figure 3b as a function of
temperature for the 5 nm thick thin film. A maximum in $R_{Hall}/R_{sheet}$ is seen to occur between 60-70 K, below
which the magnetization starts to drop in accordance with the drop in the Hall resistance as discussed earlier.
Such behavior is not seen in thick (Ga,Mn)As films, but the relative drop in magnetization with decreasing
temperature decreases with increasing (Ga,Mn)As thickness, thus indicating that this may be a thin film property.
Since the anomalous Hall term is not a direct measure of magnetization, other types of measurements are needed to
understand this behavior in more detail.

\begin{figure}
\includegraphics[width=.45\textwidth]{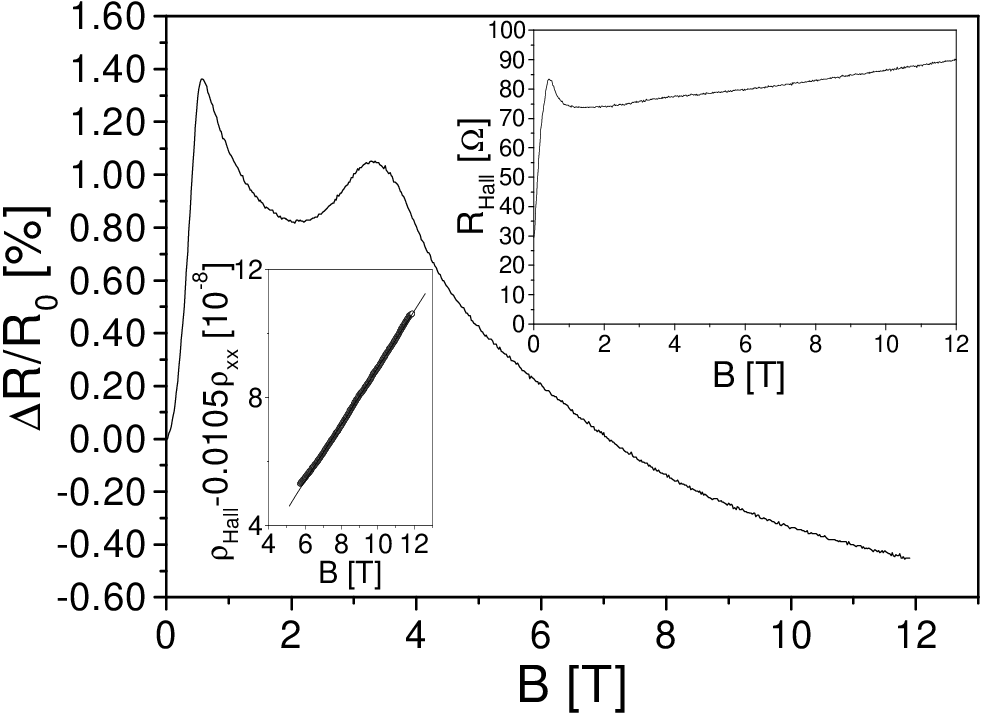}
\caption{\label{4}\ Magnetoresistance of the 5 nm thick Ga$_{0.95}$Mn$_{0.05}$As thin film at T=300 mK. The Hall
resistance and the linear fit to eq. (1), assuming Skew scattering, is shown as insets. $cM$ is the fitting
parameter.}
\end{figure}

A carrier density of ~7.1$\times$10$^{20}$ cm$^{-3}$ has been determined from Hall measurements in a magnetic field
up to 12 T for the 5 nm thick thin film at T=300 mK, by fitting the data to eq. (1). The magneto- and Hall
resistance is shown in figure 4. Beside the positive magnetoresistance at low magnetic fields a second maximum
appears at 4 T. So far we do not know to what physical mechanism we can attribute this behavior, which is not seen
in thick (Ga,Mn)As films. Due to large negative magnetoresistance, high magnetic fields are usually required in
order to determine the carrier density, but in this case, the negative magnetoresistance is less that 1$\%$ and it
is possible to obtain a linear fit from about 6 T to equation (1), assuming Skew scattering and using $cM$ as a
fitting parameter. See figure 4. The density of magnetic ions for 5$\%$ Mn is about 1.1 nm$^{-3}$. The carrier
density corresponds to an acceptor efficiency of 65$\%$, which is high compared to what is seen in thick (Ga,Mn)As
films, where the carrier density is obtained in the same way.

The RKKY approach by Boselli \emph{et al.}~\cite{cc}, though using an acceptor efficiency of 25$\%$, shows a
decrease in Curie temperature with increasing quantum well width. This is due to the increasing competition between
ferro- and antiferromagnetic exchange coupling, resulting in a non-collinear spin phase. Antiferromagnetic exchange
may play a role in these thin films, since the mean free path is determined to be 0.5 nm using an effective mass of
0.5m$_{o}$, m$_{0}$ being the electron rest mass, and the period of the RKKY oscillations is given by
$\pi/k_{f}$=1.14 nm for the 5 nm thick film. F. Matsukura \emph{et al.}~\cite{bb} relate their observed increase in
Curie temperature with decreasing (Ga,Mn)As layer thickness, to the observed dependence of magnetic anisotropy on
the thickness of the layer. The Curie temperature also depends on the carrier density~\cite{e}, but this has yet to
be investigated for these thin films.

In summary we have investigated the magnetotransport properties of very thin (Ga,Mn)As layers containing 5$\%$ Mn
in the thickness range of 3 nm to 15 nm and found the thickness limit of 5 nm for ferromagnetic phase transition in
single (Ga,Mn)As layer embedded in LT GaAs. Thinner (Ga,Mn)As films were highly resistive and show no evidence of a
para- to ferromagnetic phase transition, but the thickness limit for the FM phase in single (Ga,Mn)As may be lower
than reported here. The Curie temperature for the ferromagnetic layers is seen to decrease with increasing layer
thickness. Significant differences between the magnetotransport properties of thick and thin (Ga,Mn)As films have
been observed. Due to the very low negative magnetoresistance at 300 mK  and magnetic fields in the range of 0 - 12
T we could estimate the density of holes in 5 nm thick (Ga,Mn)As film with 5$\%$ Mn to be as high as
7.1$\times$10$^{20}$ cm$^{-3}$, which means 65$\%$ doping efficiency of Mn acceptors.

Acknowledgement: This work was supported by the Swedish Natural Science Research Council (NFR), the Swedish
Research Council for Engineering Sciences (TFR), the Nanometer Structure Consortium in Lund, the Swedish Foundation
for Strategic Research (SSF), the Danish Research Council of Engineering Sciences (STVF framework programme
"Nanomagnetism"), and  the Danish Science Research Council (SNF framework programme "Mesoscopic Physics"). The
samples are partly prepared at the III-V NANOLAB.\newline

\end{document}